\documentclass[aps,superscriptaddress,preprintnumbers,showpacs,showkeys,nofootinbib,twocolumn]{revtex4}
\usepackage{graphicx,epsfig,wrapfig,amssymb}
\newcommand{\be}{\begin{equation}}
\newcommand{\ee}{\end{equation}}
\newcommand{\ba}{\begin{eqnarray}}
\newcommand{\ea}{\end{eqnarray}}
\newcommand{\la}{\langle}
\newcommand{\ra}{\rangle}
\newcommand{\di}{ {\rm d} }

\begin{document}
\newcommand*{\Jlab}{Thomas Jefferson National Accelerator Facility,
Newport News, VA 23606, U.S.A.}\affiliation{\Jlab}
\newcommand*{\Dubna}{Joint Institute for Nuclear Research, Dubna, 
141980 Russia}\affiliation{\Dubna}
\newcommand*{\Bochum}{Institut f{\"u}r Theoretische Physik II, 
Ruhr-Universit{\"a}t Bochum, D-44780 Bochum, Germany}\affiliation{\Bochum}
\newcommand*{\Philadelphia}{Department of Physics, Barton Hall, Temple University, 
Philadelphia, PA 19122-6082, U.S.A.}

\title{Are there approximate relations among \\
transverse momentum dependent distribution functions?}
\author{H.~Avakian}\affiliation{\Jlab}
\author{A.~V.~Efremov}\affiliation{\Dubna}
\author{K.~Goeke}\affiliation{\Bochum}
\author{A.~Metz}\affiliation{\Bochum}\affiliation{\Philadelphia}
\author{P.~Schweitzer}\affiliation{\Bochum}
\author{T.~Teckentrup}\affiliation{\Bochum}
\date{September 2007}
\begin{abstract}
  Certain {\sl exact} relations among transverse momentum dependent 
  parton distribution functions due to QCD equations of motion turn
  into {\sl approximate} ones upon the neglect of pure twist-3 terms.
  On the basis of available data from HERMES we test the practical usefulness 
  of one such  ``Wandzura-Wilczek-type approximation'',
  namely of that connecting $h_{1L}^{\perp(1)a}(x)$ to $h_L^a(x)$,
  and discuss how it can be further tested by future CLAS and COMPASS data.
 \end{abstract}
\pacs{13.88.+e, 
      13.85.Ni, 
      13.60.-r, 
      13.85.Qk} 
\keywords{Semi-inclusive deep inelastic scattering, 
	  tranverse momentum dependent distribution functions, 
	  single spin asymmetry,
	  Wandzura-Wilczek approximation}
\maketitle
\section{Introduction}
\label{Sec-1:introduction}

Semi-inclusive deep inelastic lepton nucleon scattering (SIDIS), hadron 
production in electron-positron annihilations and the Drell-Yan process 
\cite{Cahn:1978se,Sivers:1989cc,Efremov:1992pe,Collins:1992kk,Collins:1993kq,Kotzinian:1994dv,Mulders:1995dh,Boer:1997nt,Boer:1997mf,Boer:1997qn,Boer:1999mm,Brodsky:2002cx,Collins:2002kn,Belitsky:2002sm} 
allow to access information on transverse momentum dependent (TMD) parton 
distribution functions (pdf) and fragmentation functions  \cite{Collins:2003fm}. 
In order to be sensitive to ``intrinsic'' transverse parton momenta it is necessary
to measure adequate transverse momenta in the final state, e.g.\  in SIDIS
the transverse momenta of produced hadrons with respect to the virtual photon.
Some data on such processes are available
\cite{Arneodo:1986cf,Airapetian:1999tv,Airapetian:2001eg,Airapetian:2002mf,Avakian:2003pk,Airapetian:2004tw,Alexakhin:2005iw,Diefenthaler:2005gx,Ageev:2006da,Avakian:2005ps,Airapetian:2005jc,Airapetian:2006rx,Abe:2005zx,Ogawa:2006bm,Martin:2007au,Diefenthaler:2007rj,Kotzinian:2007uv},
and at least in the case of twist-2 observables factorization applies 
\cite{Collins:1981uk,Ji:2004wu,Collins:2004nx}.

Eight twist-2 and sixteen twist-3 TMD pdfs describe the nucleon structure
in these processes, namely \cite{Goeke:2005hb,Bacchetta:2006tn}
\be\label{Eq:TMD-pdfs}
	\underbrace{f_1^a,\, f_{1T}^{\perp a},\, g_{1L}^a,\,g_{1T}^a,\,
   	h_{1T}^a,\, h_{1L}^{\perp a},\, h_{1T}^{\perp a},\,h_1^{\perp a},
	}_{\mbox{\footnotesize twist-2}}\,
	\underbrace{e^a,\,g_T^a,\,h_L^a,\,\dots }_{\mbox{\footnotesize twist-3}}
\ee
which are functions of $x$ and ${\bf p}_T^2$. (The dots denote thirteen 
further twist-3 TMD pdfs. The renormalization scale dependence 
is not indicated for brevity.)
Integrating over transverse momenta one is left with six independent 
``collinear'' pdfs 
\cite{Ralston:1979ys,Jaffe:1991ra}
\be\label{Eq:coll-pdfs}
	\underbrace{f_1^a(x), \;\;\; g_1^a(x), \;\;\; h_1^a(x),}_{
	\mbox{\footnotesize twist-2}} \;\;\;
	\underbrace{  e^a(x), \;\;\; g_T^a(x), \;\;\; h_L^a(x).}_{
	\mbox{\footnotesize twist-3}}
\ee
where the relations hold
$j(x) = \int\di^2{\bf p}_T j(x,{\bf p}_T^2)$ 
for $j=f_1^a,\,e^a,\,g_T,\,h_L$ 
while $g_1^a(x) = \int\di^2{\bf p}_T g_{1L}^a(x,{\bf p}_T^2)$ and
$h_1^a(x) = \int\di^2{\bf p}_T \{h_{1T}^a(x,{\bf p}_T^2)+{\bf p}_T^2/(2M_N^2)
h_{1T}^{\perp a}(x,{\bf p}_T^2)\}$.

In view of the prolification of novel functions in  (\ref{Eq:TMD-pdfs}) 
one may ask whether some of the unknown TMD pdfs could be related to 
(possibly better) known ones.
Since all structures in (\ref{Eq:TMD-pdfs}) are independent
\cite{Goeke:2005hb}, any such relations can only be approximate.

Candidates for such {\sl approximate} relations can be obtained as follows. 
From QCD equations of motion (eom), one obtains among others the following 
{\sl exact} relations \cite{Mulders:1995dh}
\ba	
   	g_{1T}^{\perp(1)a}(x)&\stackrel{\rm eom}{=}&
	x\,g_T^a(x)-x\,\tilde{g}_T^a(x) \;,\label{Eq:eom-gT}\\
    -2\,h_{1L}^{\perp(1)a}(x)&\stackrel{\rm eom}{=}&
	x\,h_L^a(x)-x\,\tilde{h}_L^a(x)
	\;,\label{Eq:eom-hL}
\ea
with the transverse moments defined as ($g_{1T}^{\perp(1)}$ analog) 
\be\label{Eq:transv-mom-hL}
	h_{1L}^{\perp (1)a}(x) \equiv \int\di^2{\bf p}_T \;
	\frac{{\bf p}_T^2}{{2M_N^2}}\;h_{1L}^{\perp a}(x,{\bf p}_T^2)\,,
\ee
and with $\tilde{g}_T^a(x)$, $\tilde{h}_L^a(x)$ denoting pure twist-3
``interaction dependent'' terms due to quark-gluon-quark correlations 
(and current quark mass terms).
In the next step, we recall the relations among the collinear 
pdfs (\ref{Eq:coll-pdfs}) \cite{Wandzura:1977qf,Jaffe:1991ra,Efremov:2002qh}
\ba\label{Eq:WW-relation-gT}
   g_T^a(x)=\int_x^1\frac{\di y}{y\;}\,g_1^a(y)+\tilde{g}_T^{\prime a}(x)\;, \\
   \label{Eq:WW-relation-hL}
   h_L^a(x)=2x\int_x^1\frac{\di y}{y^2\;}\,h_1^a(y)+\tilde{h}_L^{\prime a}(x)
   \;,\ea
where $\tilde{g}_T^{\prime a}(x)$, $\tilde{h}_L^{\prime a}(x)$ 
also denote pure twist-3 (and mass) terms \cite{Shuryak:1981pi,Jaffe:1989xx},
though different ones than in  (\ref{Eq:eom-gT},~\ref{Eq:eom-hL}).
Eqs.~(\ref{Eq:WW-relation-gT},~\ref{Eq:WW-relation-hL}) isolate 
``pure twist-3 terms'' in the ``twist-3'' pdfs $g_T^a(x)$, $h_L^a(x)$. 
This is because in 
(\ref{Eq:coll-pdfs}) the underlying ``working definition'' of twist \cite{Jaffe:1996zw}
(a pdf is ``twist $t$'' if its contribution to the cross section is suppressed, 
in addition to kinematic factors, by $1/Q^{t-2}$ with $Q$ the hard scale 
in the process) differs from the strict definition of twist 
(mass dimension of the operator minus its spin).

The remarkable observation is that $\tilde{g}_T^{\prime a}(x)$ 
is consistent with zero within error bars 
\cite{Zheng:2004ce,Amarian:2003jy,Anthony:2002hy,Abe:1998wq,Adams:1994id} 
and to a good accuracy 
\ba\label{Eq:WW-approx-gT}
	g_T^a(x) \stackrel{\rm WW}{\approx}\int_x^1\frac{\di y}{y\;}\,g_1^a(y) 
	\;\;\;\mbox{(exp.\ observation)}
\ea
which is the ``Wandzura-Wilczek (WW) approximation''.

Lattice QCD \cite{Gockeler:2000ja,Gockeler:2005vw} and the 
instanton model of the QCD vacuum \cite{Balla:1997hf} support this observation.
Interestingly the latter predicts also $\tilde{h}_L^{\prime a}(x)$ to be small 
\cite{Dressler:1999hc},  such that
\ba\label{Eq:WW-approx-hL}
	h_L^a(x) \approx 2x\int_x^1\frac{\di y}{y^2}\,h_1^a(y) 
	\;\;\;\;\mbox{(prediction).}
\ea

On the basis of this positive experimental and (or) theoretical 
experience with the smallness of pure twist-3 (and mass) terms one may suspect
that the analog terms in the relations (\ref{Eq:eom-gT},~\ref{Eq:eom-hL}) 
could also be negligible.
If true one would have valuable WW-type approximations
\ba
   	g_{1T}^{\perp(1)a}(x)&\stackrel{\rm !?}{\approx}& \phantom{-}
	x\,\int_x^1\frac{\di y}{y\;}\,g_1^a(y) \;,
	\label{Eq:WW-approx-g1T}\\
    	h_{1L}^{\perp(1)a}(x)&\stackrel{\rm !?}{\approx}& -
	x^2\!\int_x^1\frac{\di y}{y^2\;}\,h_1^a(y)\;,
	\label{Eq:WW-approx-h1L}\ea
that could be satisfied with an accuracy comparable to that of
(\ref{Eq:WW-approx-gT}). This remains to be tested in experiment. 

An immediate application (or test) for the relations 
(\ref{Eq:WW-approx-g1T},~\ref{Eq:WW-approx-h1L}) is provided
by the following single/double spin asymmetries (SSA/DSA) in SIDIS
\ba
	 A_{UL}^{\sin2\phi} &\propto&
	\sum\limits_ae_a^2\,h_{1L}^{\perp(1)a}\,H_1^{\perp a}\label{Eq:SSA-AUL2}\,,\\
     	A_{LT}^{\cos(\phi-\phi_S)} &\propto& 
	\sum\limits_ae_a^2\;g_{1T}^{\perp(1)a}\;D_1^a\,,\label{Eq:DSA-ALT}
\ea
where the first index $U$ (or $L$) means that the leptons are un- 
(or longitudinally) polarized, the second $L$ (or $T$) indicates the 
longitudinal (or transverse) polarization of the nucleon, and $\phi$ ($\phi_S$) 
denotes the azimuthal angle of the produced hadron $h$ 
(target polarization vector $S$) with respect to the axis
defined by the virtual photon, see Fig.~\ref{fig1-processes-kinematics}.
The superscripts $\sin2\phi$ or $\cos(\phi-\phi_S)$ mean that the spin asymmetries
were weighted correspondingly in order to isolate the contributions responsible for 
the particular azimuthal distributions.

In  (\ref{Eq:SSA-AUL2}) $H_1^{\perp a}$ denotes the Collins fragmentation function
\cite{Efremov:1992pe,Collins:1992kk,Collins:1993kq} on which data from SIDIS 
\cite{Airapetian:2004tw,Alexakhin:2005iw,Diefenthaler:2005gx,Ageev:2006da}
on the SSA 
\be\label{Eq:SSA-AUT}
     	A_{UT}^{\sin(\phi+\phi_S)}\propto\sum\limits_ae_a^2\,h_1^a\,H_1^{\perp a}
\ee
and from $e^+e^-$ annihilations \cite{Abe:2005zx,Ogawa:2006bm} 
give rise to a first but already consistent picture of $H_1^\perp$
\cite{Vogelsang:2005cs,Efremov:2006qm,Anselmino:2007fs}.
The 	$D_1^a$ in (\ref{Eq:DSA-ALT}) is the unpolarized fragmentation function 
which enters, of course, also the respective denominators in  
(\ref{Eq:SSA-AUL2}-\ref{Eq:SSA-AUT}) that are proportional to 
$\sum_ae_a^2\,f_1^a\,D_1^a$.

Final HERMES \cite{Airapetian:1999tv,Airapetian:2001eg,Airapetian:2002mf} and 
preliminary CLAS \cite{Avakian:2005ps} data on (\ref{Eq:SSA-AUL2}) and 
preliminary COMPASS data \cite{Kotzinian:2007uv} on  (\ref{Eq:DSA-ALT}) 
are available, such that first tests of the WW-type approximations
(\ref{Eq:WW-approx-g1T},~\ref{Eq:WW-approx-h1L}) are now or soon possible.

In this note we shall present a test of the approximation
(\ref{Eq:WW-approx-h1L}). Under the assumption that this approximation
{\sl works}, we shall see that it yields results for 
the SSA (\ref{Eq:SSA-AUL2}) compatible with HERMES data 
\cite{Airapetian:1999tv,Airapetian:2001eg,Airapetian:2002mf}.
From another point of view our work provides a first independent 
cross check from SIDIS for the emerging picture of  $H_1^\perp$
\cite{Vogelsang:2005cs,Efremov:2006qm,Anselmino:2007fs}.
The SSA (\ref{Eq:SSA-AUL2}) was recently studied in \cite{Gamberg:2007gb}.

A test of the approximation (\ref{Eq:WW-approx-g1T}) was suggested 
in \cite{Kotzinian:2006dw} along the lines of the study of the SSA 
(\ref{Eq:DSA-ALT}) discussed previously also in \cite{Kotzinian:1995cz}.

Among the eight structure functions in SIDIS described in terms of 
twist-2 pdfs and fragmentation functions \cite{Bacchetta:2006tn} 
the SSAs (\ref{Eq:SSA-AUL2},~\ref{Eq:DSA-ALT}) are the only ones, 
for which WW-type approximations could be of use.
Exact eom-relations exist, in fact, for all eight twist-2 pdfs in (\ref{Eq:TMD-pdfs}).
But the relations (\ref{Eq:eom-gT},~\ref{Eq:eom-hL}) 
are special in that they connect the respective TMD pdfs, 
namely $g_{1T}^\perp$ and $h_{1L}^\perp$, to ``collinear'' twist-3 pdfs, 
namely $g_T$ and $h_L$. Those in turn are related to twist-2 pdfs, $g_1$ and $h_1$, 
by means of (experimentally established or theoretically predicted) 
WW-approximations (\ref{Eq:WW-approx-gT},~\ref{Eq:WW-approx-hL}).

Experiments may or may not confirm that the  WW-type approximations
(\ref{Eq:WW-approx-g1T},~\ref{Eq:WW-approx-h1L}) work.

What would it mean if (\ref{Eq:WW-approx-g1T},~\ref{Eq:WW-approx-h1L}) 
were found to be satisfied to within a very good accuracy? First, that 
would be of practical use for understanding and interpreting the first data 
\cite{Airapetian:1999tv,Airapetian:2001eg,Airapetian:2002mf,Avakian:2003pk,Airapetian:2004tw,Alexakhin:2005iw,Diefenthaler:2005gx,Ageev:2006da,Avakian:2005ps,Airapetian:2005jc,Airapetian:2006rx,Abe:2005zx,Ogawa:2006bm,Martin:2007au,Diefenthaler:2007rj,Kotzinian:2007uv}.
Second, it would call for theoretical explanations why pure twist-3 terms should be
small. (Only for the smallness of the ``collinear'' pure twist-3 terms in 
(\ref{Eq:WW-approx-gT},~\ref{Eq:WW-approx-hL}) lattice QCD 
\cite{Gockeler:2000ja,Gockeler:2005vw} and/or instanton vacuum 
\cite{Balla:1997hf,Dressler:1999hc} provide explanations.)

What would it mean if (\ref{Eq:WW-approx-g1T},~\ref{Eq:WW-approx-h1L}) 
were found to work poorly? This scenario would be equally interesting.
In fact, all eight pdfs in (\ref{Eq:TMD-pdfs}) are independent structures,
and any of them contains different type of information on the internal
structure of the nucleon. The measurement of the complete set of all eighteen 
structure functions available in SIDIS \cite{Kotzinian:1994dv} is therefore 
indispensable for our aim to learn more about the nucleon structure.

One type of information accessible in this way concerns
effects related to the orbital motion of quarks, and in particular correlations 
of spin and transverse momentum of quarks which are dominated by valence quarks
and hence play a more important role at large $x$. E.g.~it
was shown that spin-orbit correlations may lead to significant 
contribution to partonic momentum and helicity distributions \cite{Avakian:2007xa} 
in large-$x$ limit.
Spin-orbit correlations are presumably of similar importance for transversity, 
and crucial for $h_{1L}^\perp$, which describes transversely polarized quarks 
in a longitudinally polarized nucleon, and is a measure for the correlation 
of the transverse spin and the transverse momentum of quarks.

This note is organized as follows.
In Sec.~\ref{Sec-2:h1Lperp-in-WW-approx} we estimate $h_{1L}^\perp$ 
by means of the WW-type approximation (\ref{Eq:WW-approx-h1L}) 
using various different models for $h_1$,
and discuss model-independent features of these estimates. 
In Sec.~\ref{Sec-3:AUL2-in-experiment} we introduce notations and definitions.
In Sec.~\ref{Sec-4:AUL2-in-theory} we evaluate the SSA (\ref{Eq:SSA-AUL2})
in the WW-type approximation (\ref{Eq:WW-approx-h1L}) and compare the results to 
available HERMES data \cite{Airapetian:1999tv,Airapetian:2001eg,Airapetian:2002mf}.
In Secs.~\ref{Sec-5:AUL2-at-CLAS} and \ref{Sec-6:AUL2-at-COMPASS} we discuss 
what can be learned from future measurements at CLAS, and COMPASS.
Sec.~\ref{Sec-7:conclusions} contains the conclusions.

\begin{figure}[b]
	\vspace{-0.7cm}
        \includegraphics[width=5.5cm]{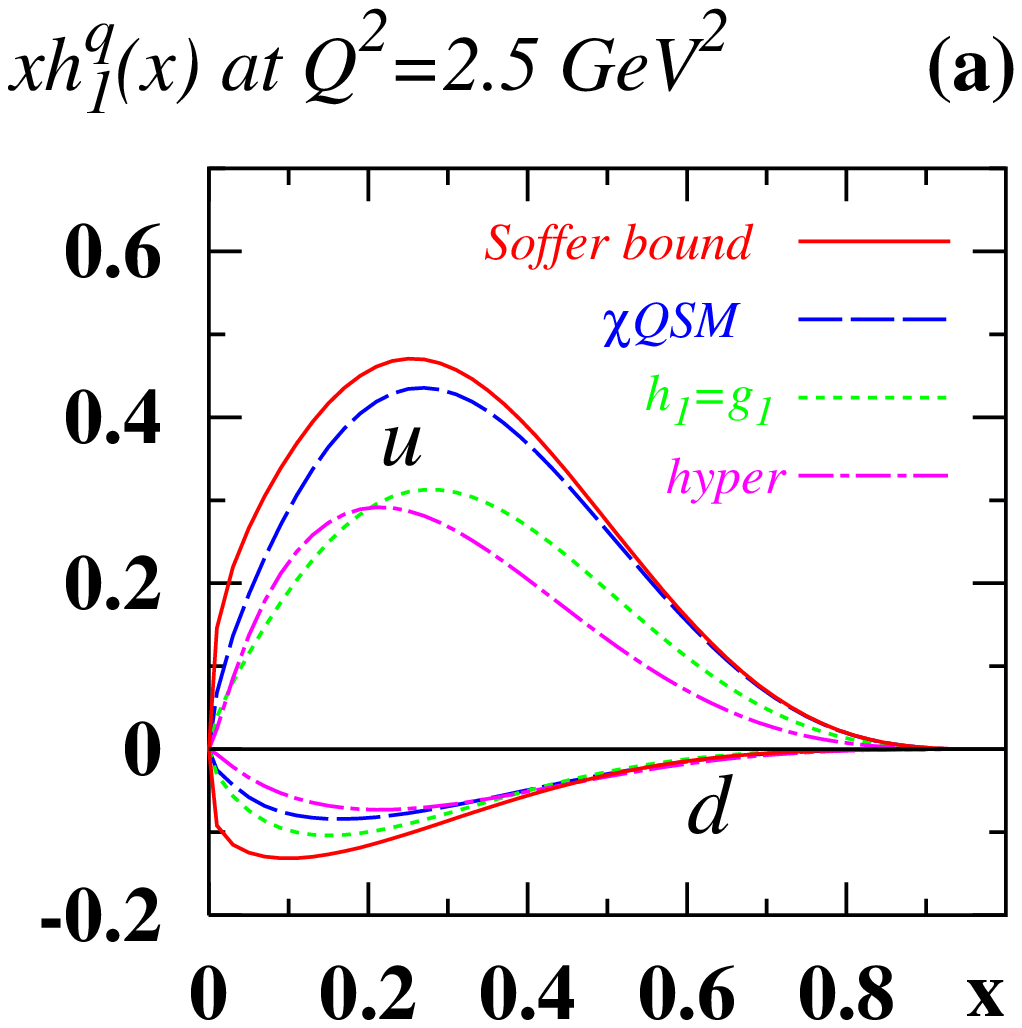}
        \includegraphics[width=5.5cm]{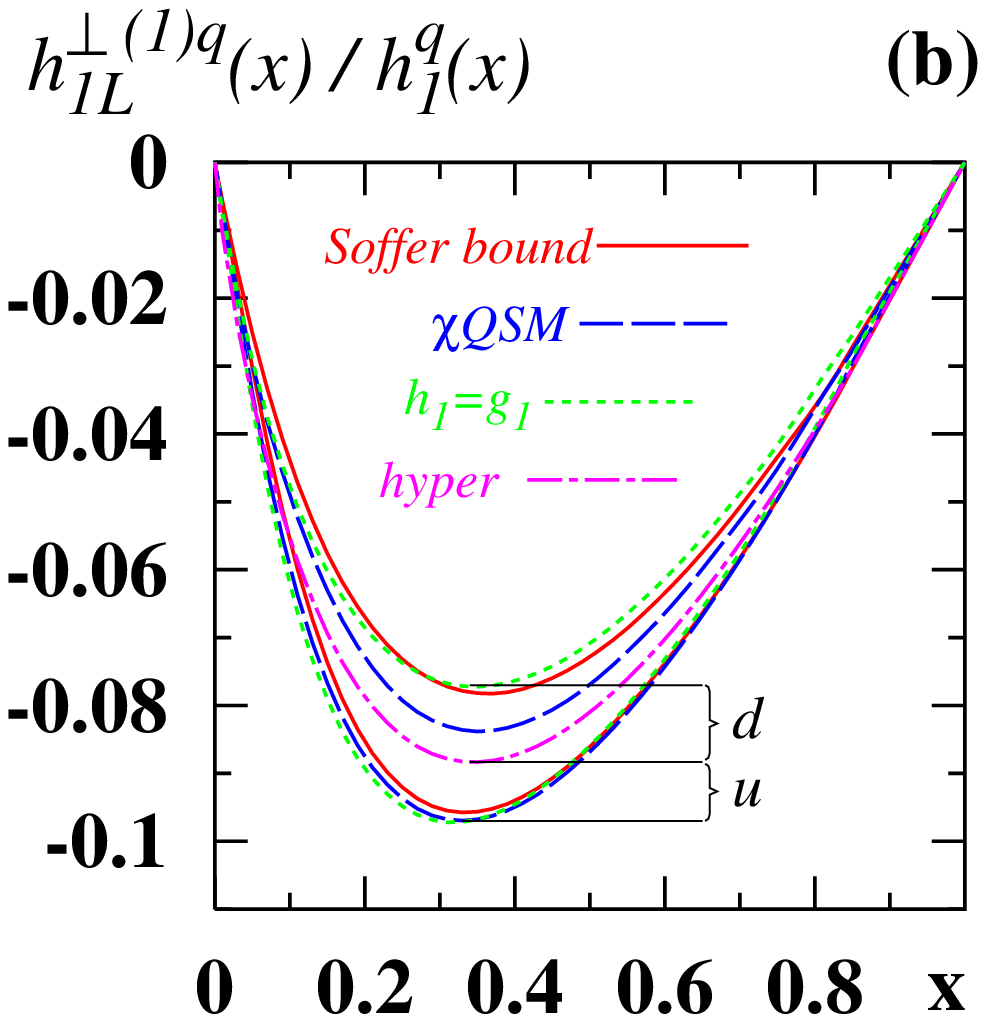}
        \includegraphics[width=5.5cm]{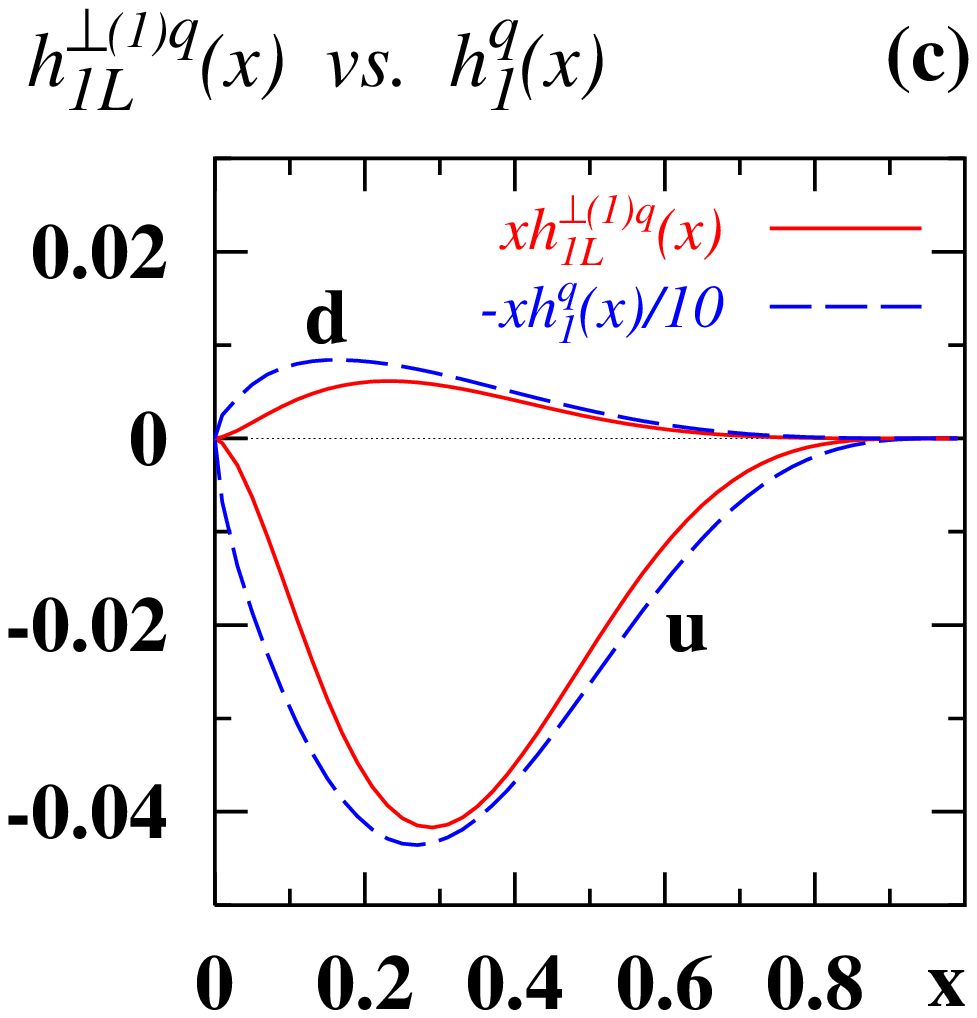}
\caption{\label{Fig01:h1Lperp-x}
	(a) Transversity, $xh_1^q(x)$, vs.\ $x$, from various models.
	(b) The ratio $h_{1L}^{\perp(1)q}(x)/h_1^q(x)$ vs.\ $x$ in 
	various models, with $h_{1L}^\perp$ estimated by means of the
	WW-type approximation (\ref{Eq:WW-approx-h1L}).\newline
	(c) $xh_{1L}^{\perp(1)q}(x)$ vs.\ $x$ from the WW-type 
	approximation (\ref{Eq:WW-approx-h1L}) and $h_1^a(x)$ from $\chi$QSM 
	\cite{Schweitzer:2001sr}, in comparison with $(-\,\frac{1}{10})xh_1^q(x)$
	from that model. All results here refer to a scale of $2.5\,{\rm GeV}^2$. }
\end{figure}

\section{\boldmath WW-type approximation for $h_{1L}^\perp$}
\label{Sec-2:h1Lperp-in-WW-approx}

In order to model $h_{1L}^{\perp(1)a}(x)$ by means of the WW-type approximation 
(\ref{Eq:WW-approx-h1L}) one inevitably has to use, in addition, models for the 
transversity pdf. Fig.~\ref{Fig01:h1Lperp-x}a shows four different models:
saturation of the Soffer bound \cite{Soffer:1994ww} at the low initial 
scale of the leading order parameterizations \cite{Gluck:1998xa,Gluck:2000dy}
(choosing $h_1^u>0$ and $h_1^d<0$),
the chiral quark soliton model ($\chi$QSM) \cite{Schweitzer:2001sr},
the non-relativistic model assumption $h_1^a(x)=g_1^a(x)$ 
at the low scale of the parameterization \cite{Gluck:2000dy},
and the hypercentral model \cite{Pasquini:2006iv}.
All curves in Fig.~\ref{Fig01:h1Lperp-x} are leading-order evolved to 
$2.5\,{\rm GeV}^2$ which is a relevant scale in experiment, see below.

These (and many other \cite{Efremov:2004tz,Barone:2001sp}) models agree on that 
$h_1^u(x)>0$ and $h_1^d(x)<0$ with $|h_1^d(x)| < h_1^u(x)$, though the predictions 
differ concerning the magnitudes, see Fig.~\ref{Fig01:h1Lperp-x}a. 
Models in which antiquark distribution functions can be computed, 
e.g.\  \cite{Schweitzer:2001sr}, predict that the transversity antiquark 
pdfs are far smaller than the quark ones.

Let us therefore establish first a robust feature of the relation 
(\ref{Eq:WW-approx-h1L}), namely the ratio $h_{1L}^{\perp(1)q}(x)/h_1^q(x)$
exhibits little dependence on the transversity model, see Fig.~\ref{Fig01:h1Lperp-x}b.

A ``universal'' behaviour of this ratio at large $x$ is not surprizing.
By inspecting (\ref{Eq:WW-approx-h1L}) for large $x$ one finds
\be\label{Eq:WW-and-large-x}
	\lim\limits_{{\rm large\,}x}\frac{h_{1L}^{\perp(1)a}(x)}{h_1^a(x)}
	\sim (1-x)\;,
\ee
which agrees with general results from large-$x$ counting rules
\cite{Brodsky:2006hj}. This is also true for (\ref{Eq:WW-approx-g1T}). 
That the WW-type approximations respect the relative large-$x$ behaviour of 
the involved pdfs can intuitively be understood by considering that 
multi-parton-correlations are likely to vanish faster at large $x$ than twist-2 terms.

Also a ``universal'' small-$x$ behaviour of the ratio can be understood
from Eq.~(\ref{Eq:WW-approx-h1L}), namely for $h_1^a(x)\sim x^\alpha$ at small $x$
one obtains
\be\label{Eq:WW-and-small-x}
	\lim\limits_{{\rm small\,}x}\frac{h_{1L}^{\perp(1)a}(x)}{h_1^a(x)}
	\sim \cases{x 		& for $\alpha\neq 1$, \cr
		    x\,\log x 	& for $\alpha=1$,}
\ee
i.e.\ the ratio tends to zero with $x\to 0$ in any case.\footnote{
	Notice that all curves in Fig.~\ref{Fig01:h1Lperp-x} are results of leading 
	order evolution \cite{Artru:1989zv} starting from low scales --- ranging 
	from $0.079\,{\rm GeV}^2$ for \cite{Pasquini:2006iv},
	till $0.36\,{\rm GeV}^2$ for $\chi$QSM \cite{Schweitzer:2001sr}. 
	Next-to-leading order evolution \cite{h1-evolution-NLO} 
	and Regge asymptotics \cite{Kirschner:1996jj} predict a 
	behaviour $h_1^a(x)\sim {\cal O}(x^0)$ for $x\to 0$.}

Nevertheless it is interesting to observe that the ratio is rather robust
also at intermediate $x$. For the hypercentral model \cite{Pasquini:2006iv} the 
ratio is flavour-independent, since there $h_1^u(x)=-4h_1^d(x)$ holds trivially due 
to the imposed SU(2)$_{\rm spin}\times$SU(2)$_{\rm flavour}$ spin-flavour-symmetry.
In the other models one, however, observes departures from that,
see Fig.~\ref{Fig01:h1Lperp-x}b.

As a common feature we finally observe
\be\label{Eq:WW-and-any-x}
	\biggl|\frac{h_{1L}^{\perp(1)a}(x)}{h_1^a(x)}\biggr| \lesssim 0.1\;.
\ee

In the following we will use the $\chi$QSM, see Fig.~\ref{Fig01:h1Lperp-x}c, 
which has several advantages. First, it is a faithful field theoretic model of 
the nucleon \cite{Diakonov:1987ty,Christov:1995vm} that describes the twist-2 pdfs 
$f_1^a(x)$ and $g_1^a(x)$ within (10-30)$\%$ accuracy \cite{Diakonov:1996sr}.
Second, this model is derived from the instanton vacuum model 
\cite{Diakonov:1983hh,Diakonov:1995qy} which predicts that the
``collinear WW-type approximation'' (\ref{Eq:WW-approx-hL})
works well \cite{Dressler:1999hc}.
Third, below we will use $h_1^a(x)$ from the $\chi$QSM in combination with information 
on the Collins effect from the analysis \cite{Efremov:2006qm} where this model was
used. This helps to minimize the model-dependence in our study.
But we shall see that our conclusions do not depend on the choice of model.

\newpage
\begin{figure}
	\vspace{1.1cm}
	\centering
        \includegraphics[width=8cm]{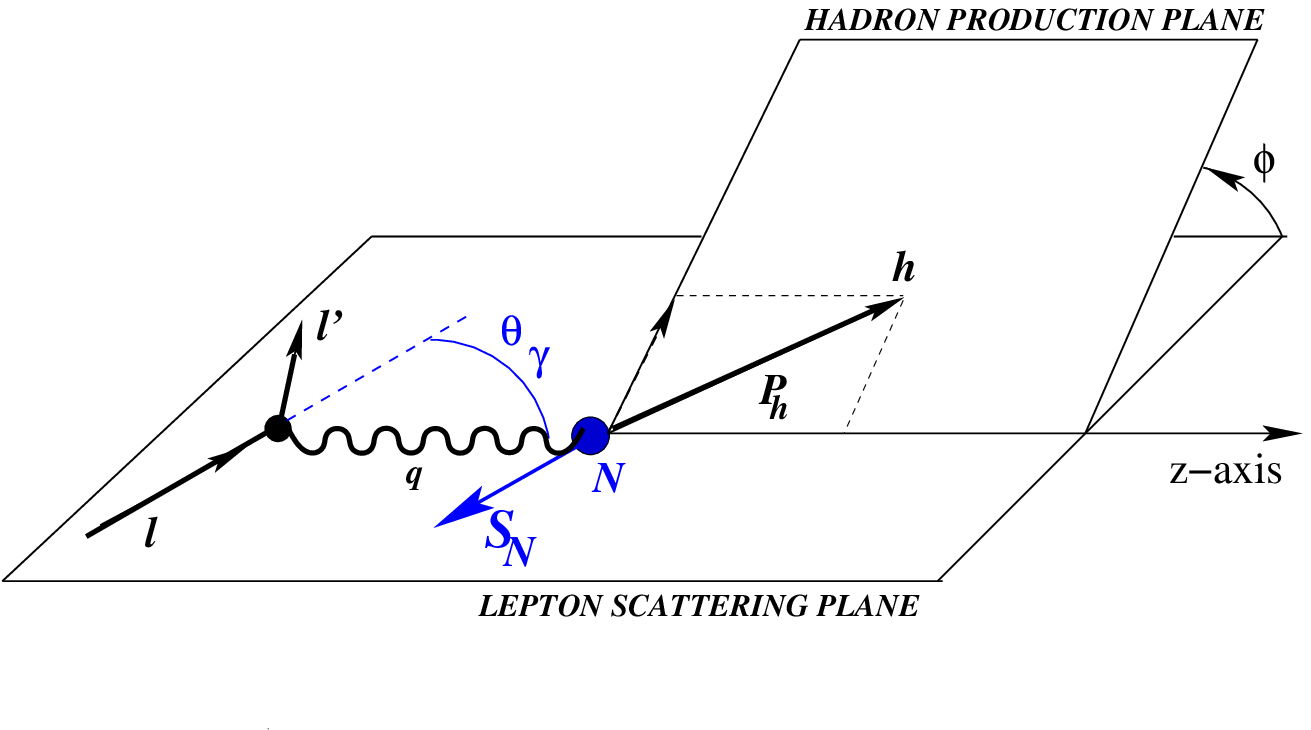}
        \caption{\label{fig1-processes-kinematics}
	Kinematics of the SIDIS process $lN\to l^\prime h X$
	and the definitions of azimuthal angles in the lab frame.	
	Here the target polarization is antiparallel to the beam (i.e.\ $\phi_S=\pi$).}
\end{figure}

\section{\boldmath $A_{UL}^{\sin2\phi}$ at HERMES}
\label{Sec-3:AUL2-in-experiment}

Let us denote the momenta of the target, incoming and outgoing lepton
by $P$, $l$ and $l'$ and introduce $s=(P+l)^2$, the four-momentum transfer 
$q= l-l'$ with $Q^2= - q^2$ and $W^2= (P+q)^2$. Then $y = Pq/Pl$ and
\be\label{notation-1}
	x = \frac{Q^2}{2Pq}	\;,\;\;\;
        z = \frac{PP_h}{Pq\;} 	\;,\;\;\; 
	\cos\theta_\gamma = 1-\frac{2M_N^2x(1-y)}{s y} \;,  \ee
where $\theta_\gamma$ denotes the angle between target polarization vector
and momentum ${\bf q}$ of the virtual photon $\gamma^\ast$,
see Fig.~\ref{fig1-processes-kinematics}, and $M_N$ is the nucleon mass. 
The component of the momentum of the produced hadron transverse with respect to 
$\gamma^\ast$ is denoted by ${\bf P}_{\!h\perp}$ and $P_{h\perp}=|{\bf P}_{\!h\perp}|$.

In the HERMES experiment $A_{UL}^{\sin2\phi}$ was measured on proton
for pion- \cite{Airapetian:1999tv,Airapetian:2001eg} and on deuteron target
for pion- and kaon-production \cite{Airapetian:2002mf} in the kinematic range
\ba\label{Eq:cuts}
&&	1\,{\rm GeV}^2 < Q^2 < 15\,{\rm GeV}^2  , \;\;\;
 	W > 2\,{\rm GeV}  			, \;\;\;\\
&&	0.023 < x < 0.4 			      \,, \;\;\;
	0.2 < y < 0.85 			      \,, \;\;\;
	0.2 <  z  < 0.7. \nonumber
\ea
The momenta of produced hadrons were subject to somehow different cuts:
$4.5\,{\rm GeV} < |{\bf P}_{\!h}| < 13.5\,{\rm GeV}$
in \cite{Airapetian:1999tv,Airapetian:2001eg}
vs.\ $2\,{\rm GeV} < |{\bf P}_{\!h}| < 15\,{\rm GeV}$ 
in \cite{Airapetian:2002mf}. The resolution cut $P_{h\perp}>50\,{\rm MeV}$ was 
applied throughout \cite{Airapetian:1999tv,Airapetian:2001eg,Airapetian:2002mf}.
This results in the following mean values
\ba
&&	\la x\ra = 0.09,\;\;\;
	\la y\ra = 0.53,\;\;\,
	\la z\ra = 0.38,\;\;\;\nonumber\\
&&	\la Q^2\ra = 2.4\,{\rm GeV}^2, \;\;\;
	\la P_{h\perp}\ra = 0.4\,{\rm GeV},\label{Eq:mean-values}\\
&&	\la Q\ra = 1.55\,{\rm GeV},\;\;\;
	\la \cos\theta_\gamma\ra = 0.98 \,.\nonumber
\ea

In the experiment the SSA was defined as
\be\label{Eq:AUL2-def-exp}
	A_{UL}^{\sin2\phi} = 
  	\frac{\sum_i\sin(2\phi_i)(N_i^{\leftrightarrows}-N_i^\rightrightarrows)}
             {\sum_i\frac12      (N_i^{\leftrightarrows}+N_i^\rightrightarrows)}
\ee
where $N_i^{\leftrightarrows}$ ($N_i^\rightrightarrows$) denotes the number
of events $i$ with target polarization antiparallel (parallel) to the beam.

\section{\boldmath $A_{UL}^{\sin2\phi}$ in WW-type approximation}
\label{Sec-4:AUL2-in-theory}

The expression for the SSA is given by \cite{Mulders:1995dh}
\be\label{Eq:AUL2-theory}
	A_{UL}^{\sin2\phi}(x) = 
	\frac{\int\di y\,[\cos\theta_\gamma(1-y)/Q^4] F_{UL}^{\sin2\phi}}
             {\int\di y\,[      (1-y+\frac12y^2)/Q^4] F_{UU,T}}
\ee
where in the notation of \cite{Bacchetta:2006tn}
the numerator is given by
\be\label{Eq:AUL2-denominator}
	F_{UU,T}(x) = \sum_ae_a^2\,xf_1^a(x)\la D_1^a\ra\,.
\ee
Since our purpose is to test the relation (\ref{Eq:WW-approx-h1L}),
we focus on the $x$-dependence of the SSA, and denote
here and in the following averages over $z$ within the
cuts (\ref{Eq:cuts}) by $\la\dots\ra=\int\di z(\dots)$.

The tree-level expression \cite{Mulders:1995dh} for the structure function 
$F_{UL}^{\sin2\phi}$ is given in terms of an 
integral which convolutes transverse parton momenta in the distribution 
and the fragmentation function 
(we neglect soft factors \cite{Ji:2004wu,Collins:2004nx}) 
\ba
	F_{UL}^{\sin2\phi}(x,z) &&\hspace{-0.3cm}= 
	\!\int\!\di^2{\bf p}_T\!\int\!\di^2{\bf K}_T \;\delta^{(2)}
	(z{\bf p}_T+{\bf K}_T-{\bf P}_{\!h\perp})\nonumber\\
&&	\hspace{-0.3cm}\times\biggl[
	\frac{2({\bf e}_h{\bf p}_T)({\bf e}_h{\bf K}_T)-({\bf p}_T{\bf K}_T)}{
	      M_Nm_h}\biggr]	\nonumber\\
&& 	\hspace{-0.3cm}\times\sum_ae_a^2\,
	xh_{1L}^{\perp a}(x,{\bf p}_T^2)
	\frac{H_1^{\perp a}(z,{\bf K}_T^2)}{z}\,,\label{Eq:AUL2-numerator}
\ea
where ${\bf e}_h={\bf P}_{\!h\perp}/P_{h\perp}$ and $m_h$ denotes the mass of 
the produced hadron.

Had the events in the numerator of (\ref{Eq:AUL2-def-exp}) 
been weighted by $P_{h\perp}^2/(M_N m_h)$ in addition to $\sin(2\phi)$,
the convolution integral could be solved in a model independent way with
the result given in terms of the transverse moment (\ref{Eq:transv-mom-hL})
of $h_{1L}^\perp$ and an analog moment for $H_1^\perp$ \cite{Boer:1997nt}.
Including such an additional weight makes data analysis more difficult due to 
acceptance effects. Omitting it, however, forces one to resort to models.

We shall assume the distributions of transverse parton momenta to be Gaussian
(and the respective widths $\la{\bf p}_{h_{1L}}^2\ra$ and 
$\la{\bf K}_{H_1}^2\ra$
to be flavour and $x$- or $z$-independent):
\ba\label{Eq:Gauss-ansatz}
   	h_{1L}^{\perp a}(x,{\bf p}_T^2) & \equiv & h_{1L}^{\perp a}(x)\; 
   \frac{\exp(-{\bf p}_T^2/\la{\bf p}^2_{h_{1L}}\ra)}{\pi\;\la{\bf p}^2_{h_{1L}}\ra}
	\;,\nonumber\\
	H_1^{\perp a}(z,{\bf K}_T^2) & \equiv & 
	H_1^{\perp a}(z)\;
	\frac{\exp(-{\bf K}_T^2/\la{\bf K}^2_{H_1}\ra)}{\pi\;\la{\bf K}^2_{H_1}\ra}
	\;.\ea
The normalizations are such that one obtains for the unpolarized functions 
$f_1^a(x)=\int\di^2{\bf p}_T\,f_1^a(x,{\bf p}_T)$
and $D_1^a(z)=\int\di^2{\bf K}_T\,D_1^a(z,{\bf K}_T)$ with analog Ans\"atze.

The Gauss Ansatz satisfactorily describes data on many hard reactions 
\cite{D'Alesio:2004up}, provided the transverse momenta are much smaller 
than the hard scale of the process, i.e.\  $\la P_{h\perp}\ra\ll \la Q\ra$
which is the case at HERMES, see (\ref{Eq:mean-values}).
In fact, the $z$-dependence of $\la P_{h\perp}\ra$ at HERMES \cite{Airapetian:2002mf}
is well described in the Gauss Ansatz \cite{Collins:2005ie}.

Of course, one has to keep in mind that (\ref{Eq:Gauss-ansatz}) 
is a crude approximation, and it is not clear whether it works 
also for polarized distribution and fragmentation functions.
Moreover, since also unintegrated forms of (\ref{Eq:eom-gT},~\ref{Eq:eom-hL}) hold, 
this Ansatz cannot be equally valid for all pdfs.

What is convenient for our purposes is that (\ref{Eq:Gauss-ansatz}) 
allows to solve the convolution integral (\ref{Eq:AUL2-numerator}). 
We obtain
\be\label{Eq:AUL2-numerator-II}
	F_{UL}^{\sin2\phi}(x) = \sum_ae_a^2\,xh_{1L}^{\perp(1)a}(x)
	\la C_{\rm Gauss} H_1^{\perp(1/2)a} \ra \;.
\ee
The  $1/2$-transverse-moment $H_1^{\perp(1/2)a}(z)$ and $C_{\rm Gauss}(z)$,
which is also a function of the Gauss model parameters, are defined
in App.~\ref{App:details}. On the basis of the 
information on the Collins effect from the 
analyses \cite{Vogelsang:2005cs,Efremov:2006qm,Anselmino:2007fs}
we estimate
\ba\label{Eq:C-Gauss-H1perp12-fav}
   	\la C_{\rm Gauss}H_1^{\perp(1/2)\rm fav}\ra 
	\approx  \;\;\;	 (0.035\pm 0.008)\times(2.2^{+2.1}_{-0.1})\;, && \;\;\\
   \label{Eq:C-Gauss-H1perp12-unf}
   	\la C_{\rm Gauss}H_1^{\perp(1/2)\rm unf}\ra
	\approx 	-(0.038\pm 0.007)\times(2.2^{+2.1}_{-0.1})\;. && \;\;
\ea
The first factors, with errors due to statistical accuracy of the 
(preliminary) HERMES data \cite{Diefenthaler:2005gx}, are from \cite{Efremov:2006qm}. 
The second factors are due to the transverse momentum dependence of the Collins 
function; their sizeable uncertainties reflect that the latter is presently 
poorly constrained by data \cite{Anselmino:2007fs}.
See App.~\ref{App:details} for details.

The errors in (\ref{Eq:C-Gauss-H1perp12-fav},~\ref{Eq:C-Gauss-H1perp12-unf})
are estimated conservatively, such that deviations 
from our predictions for the SSA should be attributed 
alone to the failure of (\ref{Eq:WW-approx-h1L}).

For the estimate of $h_{1L}^{\perp(1)a}(x)$ by means of (\ref{Eq:WW-approx-h1L}) 
we use predictions for the chiral quark-soliton model for $h_1^a(x)$ 
\cite{Schweitzer:2001sr} as shown in Fig.~\ref{Fig01:h1Lperp-x}c,
see Sec.~\ref{Sec-2:h1Lperp-in-WW-approx}.

%
\begin{figure}[t!]
\vspace{-0.2cm}
\begin{tabular}{cc}
    \hspace{-0.5cm}{\epsfxsize=1.85in\epsfbox{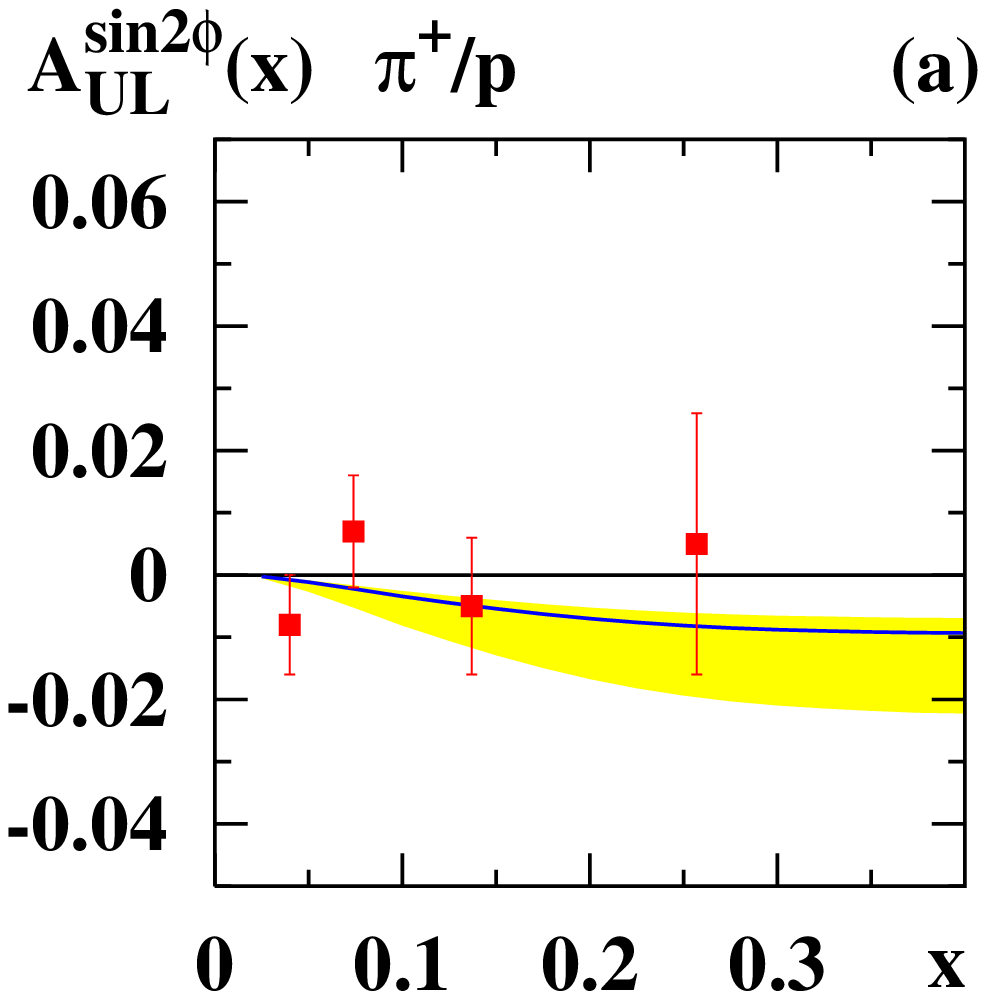}} &
    \hspace{-0.5cm}{\epsfxsize=1.85in\epsfbox{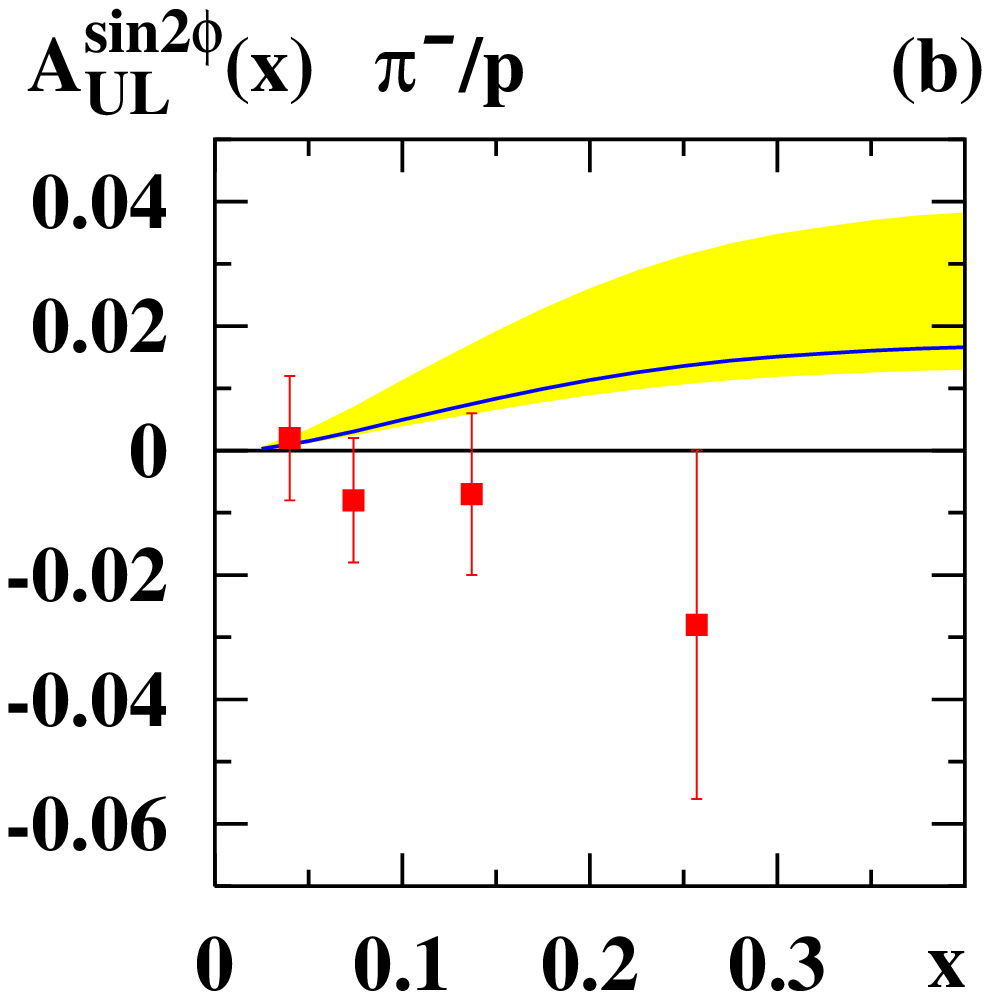}} \\
    \hspace{-0.5cm}{\epsfxsize=1.85in\epsfbox{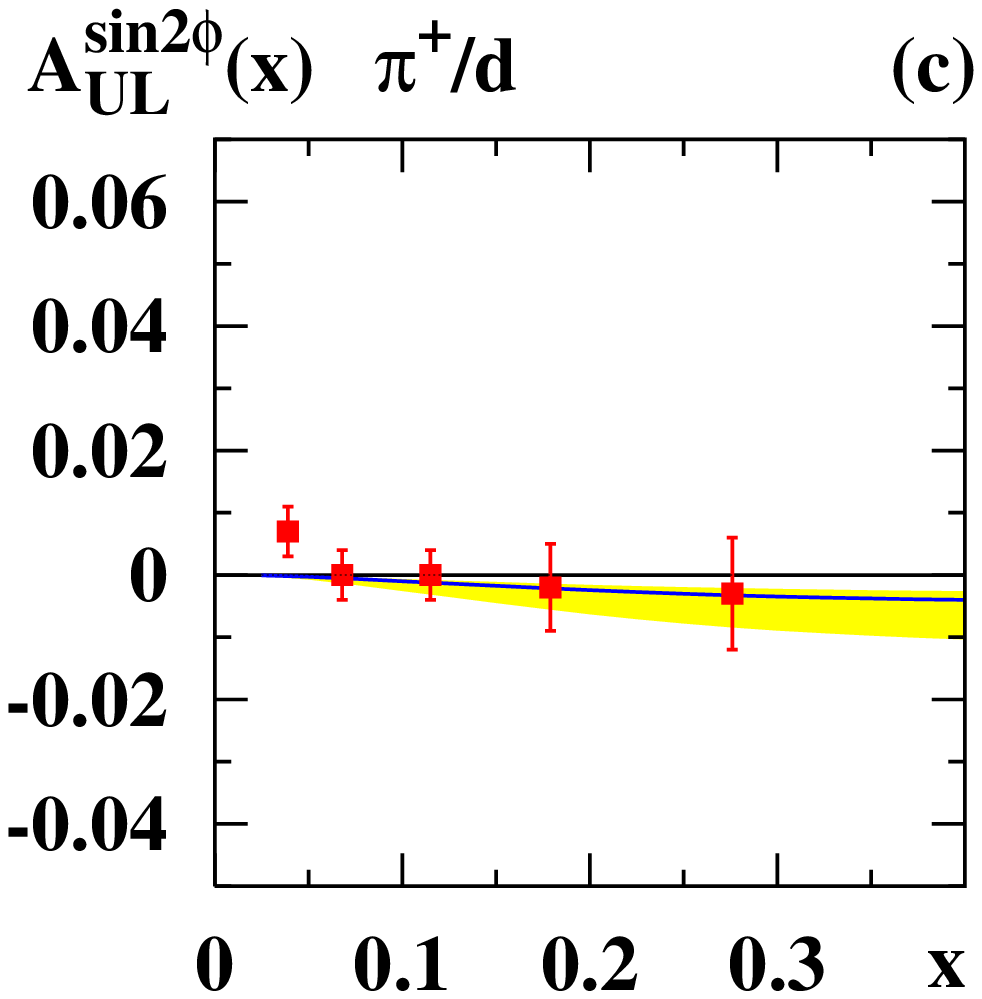}} &
    \hspace{-0.5cm}{\epsfxsize=1.85in\epsfbox{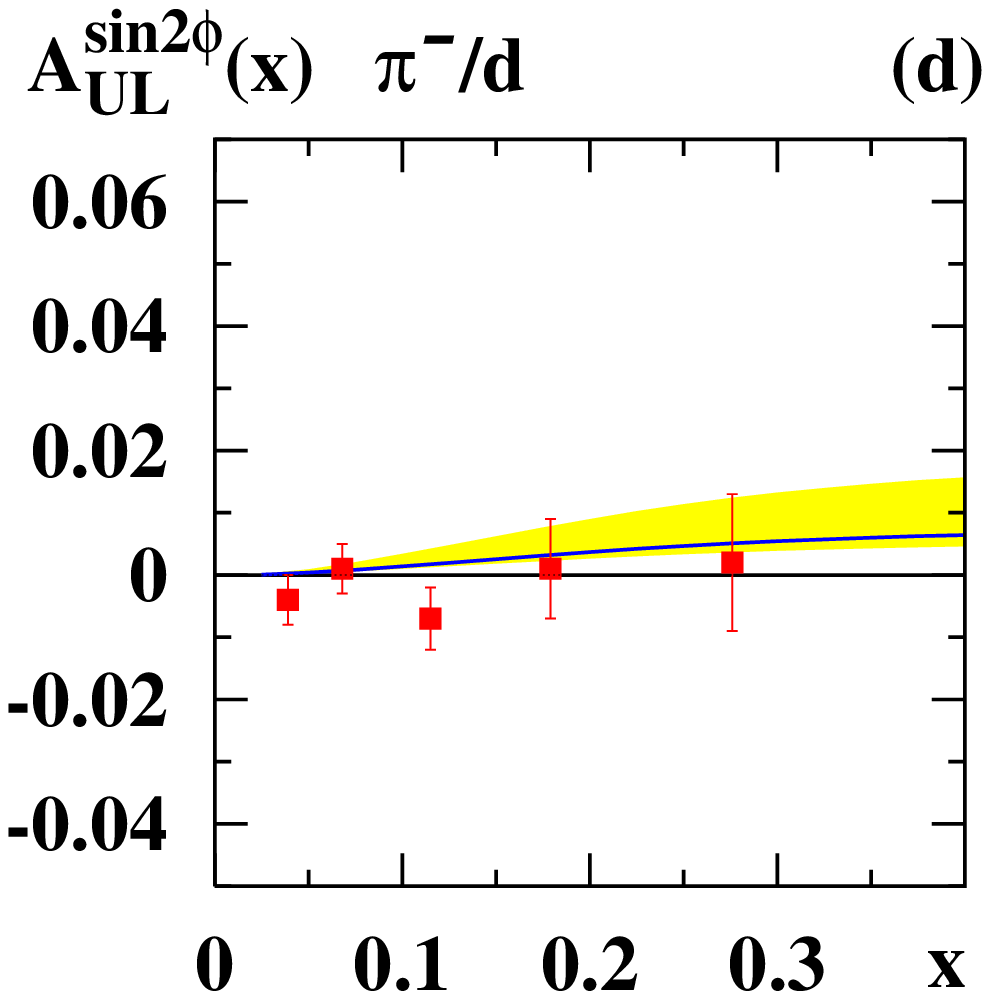}} \\
    \hspace{-0.5cm}{\epsfxsize=1.85in\epsfbox{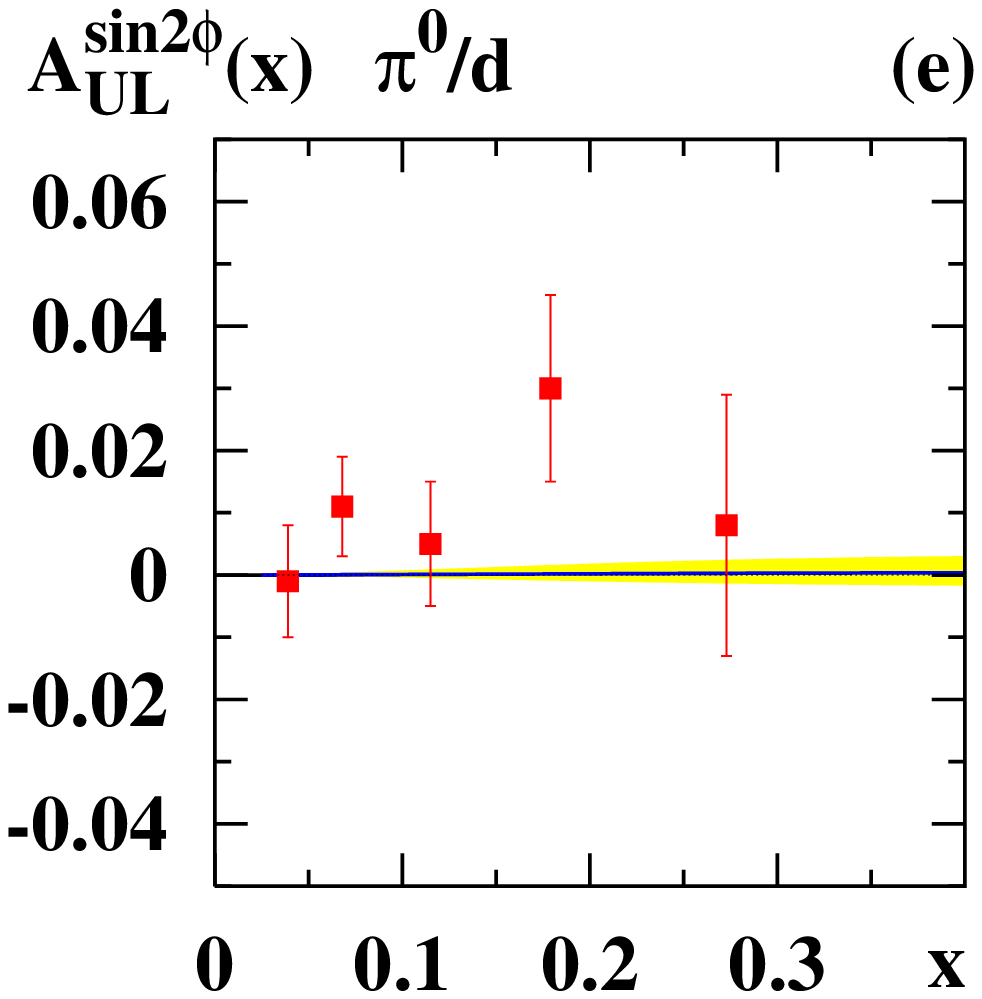}} &
    \hspace{-0.5cm}{\epsfxsize=1.85in\epsfbox{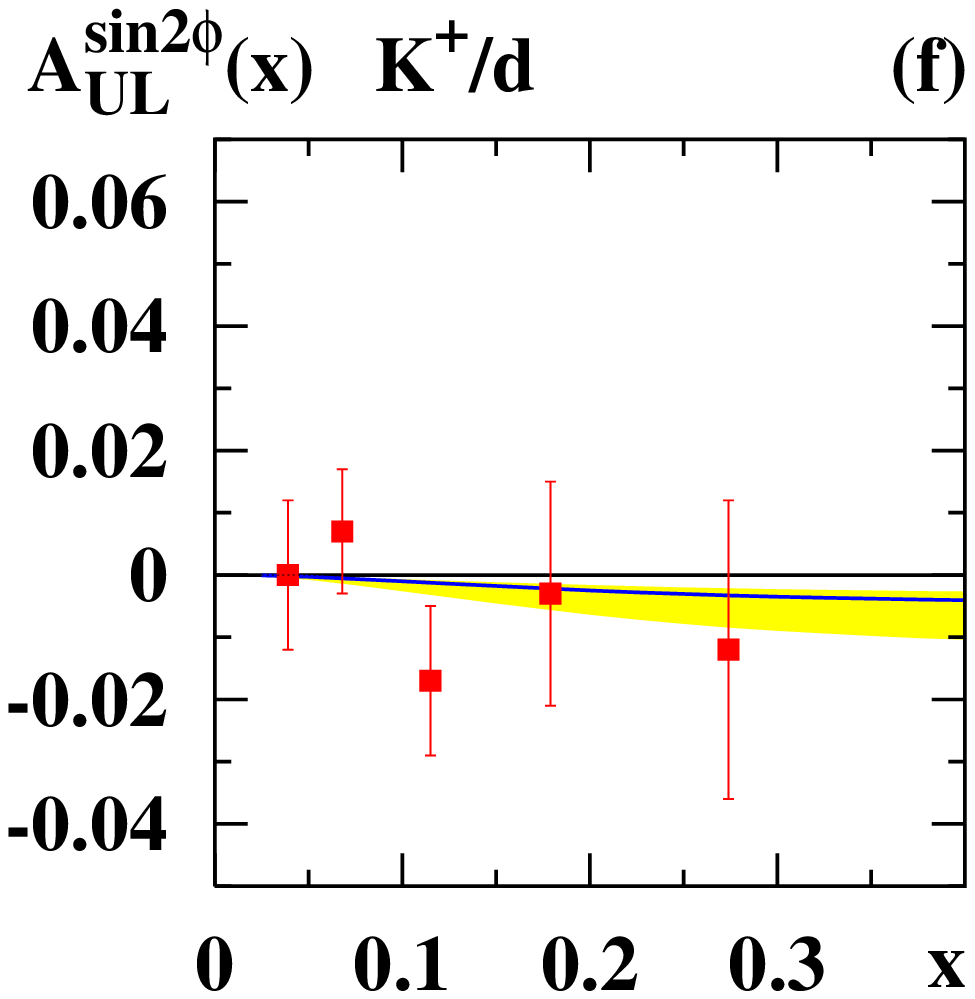}}  
\end{tabular}
\caption{\label{Fig03:AUL2-x}
 	Longitudinal target SSA $A_{UL}^{\sin2\phi}$ as function of~$x$.
	The proton (a, b) and deuterium (c-f) target data  
        are from HERMES \cite{Airapetian:1999tv,Airapetian:2002mf}.
	The theoretical curves are obtained using information on the Collins 
        fragmentation function from \cite{Efremov:2006qm,Anselmino:2007fs},
        predictions from the instanton vacuum model and chiral quark soliton 
	model for $h_L^a(x)$ and $h_1^a(x)$ \cite{Dressler:1999hc,Schweitzer:2001sr},
        and --- this is crucial in our context --- assuming 
	the validity of the WW-type approximation (\ref{Eq:WW-approx-h1L}).
	The shaded error bands are due to the uncertainties in 
	(\ref{Eq:C-Gauss-H1perp12-fav},~\ref{Eq:C-Gauss-H1perp12-unf}), 
	see App.~\ref{App:details} for details.}
\end{figure}
%

Our results shown in Figs.~\ref{Fig03:AUL2-x}a-e for pion production 
from proton and deuteron targets are consistent with the HERMES data
\cite{Airapetian:1999tv,Airapetian:2001eg,Airapetian:2002mf}, and 
do not exclude that (\ref{Eq:WW-approx-h1L}) is a useful approximation.

Fig.~\ref{Fig03:AUL2-x}f shows also the SSA for $K^+$ production.
Also here our result is compatible with data \cite{Airapetian:2002mf},
however, in this case one tests in addition assumptions on the kaon 
Collins effect, see App.~\ref{App-B:kaon-Collins-effect}.

\section{\boldmath $A_{UL}^{\sin2\phi}$ at CLAS}
\label{Sec-5:AUL2-at-CLAS}

One may roughly expect 
$|A_{UL}^{\sin2\phi}|\lesssim\frac15|A_{UT}^{\sin(\phi-\phi_S)}|$ on the 
basis of the approximation (\ref{Eq:WW-approx-h1L}), see App.~\ref{App:details}. 
Thus, $A_{UL}^{\sin2\phi}$ could be far more difficult to measure than the 
transverse target Collins effect SSA. 
Therefore what is needed is a high luminosity experiment sensitive to the region 
$0.2\lesssim x\lesssim 0.5$, where the suppression of $h_{1L}^{\perp(1)a}$ 
with respect to $h_1^a(x)$ is less pronounced.

Higher statistics at CLAS at Jefferson Lab, due to two orders of magnitude higher 
luminosity, provides access to much larger $x$ and larger $z$ than HERMES and COMPASS.
Large $z$ may also enhance the SSA due to Collins function
$H_1^{\perp(1/2)a}(z) \propto zD_1^a(z)$, as  observed in \cite{Efremov:2006qm}.
This makes CLAS an ideal experiment for studies of this SSA in particular and
spin-orbit correlations in general.
Comparison of the various data sets will also allow to draw 
valuable conclusions on the energy dependence of the process, 
possible power-corrections, etc.

The preliminary data from CLAS \cite{Avakian:2005ps} have shown non-zero SSAs
for charged pions, and a compatible with zero within error bars result for $\pi^0$. 
Within our approach it is possible to understand the results for $\pi^+$ and $\pi^0$,
however, we obtain for $\pi^-$ an opposite sign compared to the data.
In view of this observation, it is worth to look again
on Fig.~\ref{Fig03:AUL2-x}b which shows HERMES data on the $\pi^-$-SSA.
Does Fig.~\ref{Fig03:AUL2-x}b hint at an incompatibility?
Charged pions and in particular the $\pi^-$ may have significant higher twist 
contributions, in particular from exclusive vector mesons and semi-exclusive pion 
production at large $z$. 

New data expected from CLAS with $E_{\rm beam}=6\,{\rm GeV}$ \cite{avak-clas6}, will 
increase the existing statistics by about an order of magnitude and more importantly
provide comparable to $\pi^+$ sample of $\pi^0$ events. Neutral pion sample 
is not expected to have any significant contribution from exclusive vector mesons, 
neither it is expected to have significant higher twist corrections due to 
semi-exclusive production 
of pions with large $z$ \cite{Afanasev:1996mj}, where the separation between target
and current fragmentation is more pronounced.  

Higher statistics of upcoming CLAS runs at 6 \cite{avak-clas6} and 12 GeV 
\cite{avak-clas12}
will provide access also to higher values of $Q^2$ where contributions from 
exclusive and 
semi-exclusive processes are more suppressed.

JLab upgrade to 12 GeV will allow to run at an order of magnitude
higher luminosities than current CLAS, providing a comprehensive set of single and 
double spin asymmetries covering a wide range in $x$ and $z$.
That will allow detailed studies
of kinematic dependences of target SSA and clarify the situation.

\section{\boldmath $A_{UL}^{\sin2\phi}$ at COMPASS}
\label{Sec-6:AUL2-at-COMPASS}

COMPASS has taken data with a longitudinally polarized deuterium target 
which are being analyzed. In near future also a proton target will be used.
The $160\,{\rm GeV}$ muon beam available at COMPASS
allows to extend the measurements of $A_{UL}^{\sin2\phi}$ and other SSAs 
into the small $x$-region. By combining all data for $Q^2> 1\,{\rm GeV}^2$
the average $\la Q^2\ra $ at COMPASS is comparable to that at HERMES.
Therefore, Figs.~\ref{Fig03:AUL2-x}a--d show roughly our predictions
for COMPASS for charged hadron production (at COMPASS about $90\%$ of the 
produced charged hadrons are pions).

From (\ref{Eq:WW-and-small-x},~\ref{Eq:WW-and-any-x}) one may expect  
$A_{UL}^{\sin2\phi}$ to be substantially smaller, especially at small $x$,
than the transverse target SSA $A_{UT}^{\sin(\phi+\phi_S)}$ found compatible 
with zero in the COMPASS deuterium target experiment 
\cite{Alexakhin:2005iw,Ageev:2006da,Martin:2007au}.

It will be interesting to see whether these predictions will be 
confirmed by COMPASS.

\section{Conclusions}
\label{Sec-7:conclusions}

The longitudinal SSA 
\cite{Airapetian:1999tv,Airapetian:2001eg,Airapetian:2002mf,Avakian:2003pk}
were subject to intensive, early studies 
\cite{DeSanctis:2000fh,Anselmino:2000mb,Efremov:2001cz,Efremov:2001ia,Ma:2002ns}
that were based on assumptions concerning the flavour dependence of $H_1^\perp$ 
\cite{Bacchetta:2002tk,Efremov:2003eq,Efremov:2004ph} that 
turned out not to be supported by data on Collins effect from SIDIS
with transverse target polarization
\cite{Airapetian:2004tw,Alexakhin:2005iw,Diefenthaler:2005gx,Ageev:2006da}
and $e^+e^-$-annihilations \cite{Abe:2005zx,Ogawa:2006bm}. 
These data  give rise to a new, consistent picture of 
$H_1^\perp$ \cite{Vogelsang:2005cs,Efremov:2006qm,Anselmino:2007fs}
which invites reanalyses of longitudinal SSA.

In this work we did this for 
$A_{UL}^{\sin2\phi} \propto \sum_ae_a^2h_{1L}^{\perp(1)a}H_1^{\perp a}$ 
from the particular point of view of the question whether 
there are useful, approximate relations among different TMD pdfs.
In fact, QCD equations of motion relate the pdf entering this SSA
to $h_L^a(x)$ and certain pure twist-3 (and quark mass) terms.
Neglecting such terms yields an approximation for $h_{1L}^{\perp(1)a}$
similar in spirit to the WW-approximation for $g_T^a(x)$ 
that is supported by data.

Our study reveals that data do not exclude the possibility that such
WW-type approximations work.
As a byproduct we observe that data on the two SSAs due to
Collins effect, $A_{UL}^{\sin2\phi}$ and $A_{UT}^{\sin(\phi+\phi_S)}$,
are compatible.

In Ref.~\cite{Kotzinian:2006dw} predictions for
$A_{LT}^{\cos(\phi-\phi_S)} \propto \sum_ae_a^2g_{1T}^{(1)a}D_1^a$ 
were made assuming the validity of a WW-type approximation for the relevant pdf.
Comparing these predictions to preliminary COMPASS data \cite{Kotzinian:2007uv}
one arrives at the same conclusion. Also here data do not exclude the possibility
that the WW-type approximation works.

In order to make more definite statements precise measurements of these
SSAs are necessary, preferably in the region around $x\sim 0.3 $ where 
the SSAs are largest. 
An order of magnitude more data on target SSA expected from CLAS upcoming run 
\cite{avak-clas6} will certainly improve our current understanding of 
this and other SSAs and shed light on spin-orbit correlations.

The value of a precise $A_{UL}^{\sin2\phi}$ should not be underestimated.
This SSA is in any case an independent source of information on the Collins 
effect. An experimental confirmation of the utility of the WW-type 
approximation (\ref{Eq:WW-approx-h1L}), however, would mean that it is 
possible to extract information on transversity, via (\ref{Eq:WW-approx-h1L}),
from a longitudinally polarized target.

\vspace{0.5cm}

\noindent{\bf Acknowledgements.}
We thank 
M.~Anselmino and A.~Prokudin for valuable discussions.
The work is partially supported by BMBF (Verbundforschung), 
German-Russian collaboration (DFG-RFFI) under contract number 436 RUS 113/881/0
and is part of the European Integrated Infrastructure Initiative Hadron
Physics project under contract number RII3-CT-2004-506078. 
A.~E.~is also supported by the Grants RFBR 06-02-16215 and 07-02-91557, 
RF MSE RNP.2.2.2.2.6546 (MIREA) and by the Heisenberg-Landau Program of JINR.
The work was supported in part by DOE contract DE-AC05-06OR23177, under
which Jefferson Science Associates, LLC,  operates the Jefferson Lab.

\appendix
\section{Pion Collins effect}
\label{App:details}

Within the Gauss model one can, of course, rewrite the expression for
the SSA (\ref{Eq:SSA-AUL2}) in many ways. However, we are interested
in exploring the approximation (\ref{Eq:WW-approx-h1L}) and wish 
to introduce the transverse moment (\ref{Eq:transv-mom-hL}) of 
$h_{1L}^{\perp a}$ which in the Gauss model is given by
\be\label{App-Eq:transv-mom-hL-Gauss}
	h_{1L}^{\perp(1)a}(x) \stackrel{\rm Gauss}{=} 
	\frac{\la{\bf p}^2_{h_{1L}}\ra}{2M_N^2}\,h_{1L}^{\perp   a}(x)\;.
\ee
In order to use information on the Collins function from the analysis 
of HERMES data \cite{Diefenthaler:2005gx} in Ref.~\cite{Efremov:2006qm}
(the reasons why here this is preferable, are explained in 
Sec.~\ref{Sec-4:AUL2-in-theory}) we introduce the (1/2)-transverse moment of 
$H_1^\perp$ which is defined as and given in Gauss model by
\ba
	H_1^{\perp(1/2)a}(z)
	&\equiv&  
	\int\!\!\di^2{\bf K}_T\,
	\frac{|{\bf K}_T|}{2zm_\pi}\,H_1^{\perp a}(z,{\bf K}_T) \nonumber\\
	&\stackrel{\rm Gauss}{=}&
	\frac{\sqrt{\pi}\la{\bf K}_{H_1}^2\ra^{1/2}}{4m_\pi\,z}\,H_1^{\perp a}(z) \;.
	\label{App-Eq:H1perp-1/2-mom}
\ea

With the above definitions the numerator of $A_{UL}^{\sin2\phi}$ is given by
(\ref{Eq:AUL2-numerator-II}) with the function $C_{\rm Gauss}$ defined as
\be\label{App-Eq:apower-01}
	C_{\rm Gauss}(z) = 
	\frac{8zM_N}{(\pi\la{\bf K}_{H_1}^2\ra)^{1/2}\,}\;
	\frac{1}{1+z^2\la{\bf p}^2_{h_{1L}}\ra/\la{\bf K}^2_{H_1}\ra}\;.
\ee
In \cite{Efremov:2006qm} the following information on the Collins effect 
was obtained from HERMES data \cite{Diefenthaler:2005gx} on the SSA 
(\ref{Eq:SSA-AUT}): 
\ba\label{App-Eq:B-Gauss-H1perp12-fav}
   &&	\la 2B_{\rm Gauss}H_1^{\perp(1/2)\rm fav}\ra =\;\;\;(3.5\pm 0.8)\% \;, \\
   \label{App-Eq:B-Gauss-H1perp12-unf}
   &&	\la 2B_{\rm Gauss}H_1^{\perp(1/2)\rm unf}\ra = -(3.8\pm 0.7)\% \;,  
\ea
with
\be
	B_{\rm Gauss}=
	\frac{1}{\sqrt{1+z^2\la{\bf p}^2_{h_1}\ra/\la{\bf K}^2_{H_1}\ra}}\;,
\ee
where $\la{\bf p}^2_{h_1}\ra$ is the Gaussian width of the transversity pdf.

In order to use the results 
(\ref{App-Eq:B-Gauss-H1perp12-fav},~\ref{App-Eq:B-Gauss-H1perp12-unf}) 
we approximate
\be\label{App-Eq:C-Gauss-approx}
	\!\!\!\la C_{\rm Gauss} H_1^{\perp(1/2)a} \ra 
	\!\approx\! \frac{4\la z\ra M_N}{(\pi\la{\bf K}_{H_1}^2\ra)^{1/2}} 
	\;\underbrace{\!\!\biggl\la\!
	\frac{2H_1^{\perp(1/2)a}}{1+z^2\la{\bf p}^2_{h_{1L}}\ra/
	\la{\bf K}^2_{H_1}\ra}\!
	\biggr\ra\!\!}_{\displaystyle \approx
	\la 2B_{\rm Gauss}H_1^{\perp(1/2)a}\ra}\,.
\ee
For $\la{\bf K}_{H_1}^2\ra$ we use results from \cite{Anselmino:2007fs} where 
Collins function was also assumed to exhibit a Gaussian $k_T$-dependence.
In the notation of \cite{Anselmino:2007fs} one has
\be
	\frac{1}{\la{\bf K}^2_{H_1}\ra} 
	= \frac{1}{\la{\bf K}^2_{D_1}\ra}
	+ \frac{1}{M^2}
\ee
where the width of the unpolarized fragmentation function was fixed from a
study of data on the Cahn effect \cite{Anselmino:2005nn} 
$\la{\bf K}^2_{D_1}\ra = 0.20\,{\rm GeV}^2$. The parameter $M$ was fitted
to data from SIDIS and $e^+e^-$-annihilations (neglecting evolution effects)
to be $M^2=(0.70 \pm 0.65)\,{\rm GeV}^2$ \cite{Anselmino:2007fs}.
This yields for the first factor in Eq.~(\ref{App-Eq:C-Gauss-approx}) 
\be\label{App-Eq:C-Gauss-approx-fin}
	\frac{4M_N\la z\ra}{(\pi\la{\bf K}^2_{H_1}\ra)^{1/2}} \simeq
	2.2^{+2.1}_{-0.1}\;.
\ee
Using for $f_1^a(x)$ and $D_1^a(z)$ the LO parameterizations 
\cite{Gluck:1998xa,Kretzer:2001pz} at $Q^2=2.5\,{\rm GeV}^2$
gives the results in Fig.~\ref{Fig03:AUL2-x}.

A remark concerning the error estimates in Fig.~\ref{Fig03:AUL2-x} 
is in order. Strictly speaking the errors in
 (\ref{App-Eq:B-Gauss-H1perp12-fav},~\ref{App-Eq:B-Gauss-H1perp12-unf}) 
and (\ref{App-Eq:C-Gauss-approx-fin}) are not independent but correlated which 
we disregard. This means that the errors in Fig.~\ref{Fig03:AUL2-x} are somehow 
overestimated.
In view of the approximations we make, however, this is not undesired,
as it helps to estimate the errors more conservatively.
With such more conservative error estimates we are on the safe side
from the point of view of testing the WW-type approximation (\ref{Eq:WW-approx-h1L}).
In fact, a deviation of our results from data would then presumably
be due to a failure of the approximation (\ref{Eq:WW-approx-h1L}).

We notice the following rough estimate.
From (\ref{Eq:WW-and-any-x}) and the mean value in 
(\ref{App-Eq:C-Gauss-approx-fin}) one may estimate roughly
\be\label{Eq:AUL2-vs-AUT}
	|A_{UL}^{\sin2\phi}|\lesssim\frac15|A_{UT}^{\sin(\phi-\phi_S)}|\,,
\ee
as other factors in the two SSAs are either the same or of similar magnitude.

\vspace{0.3cm}

\section{Kaon Collins effect}
\label{App-B:kaon-Collins-effect}

We also wish to estimate the SSA for $K^+$. For that we notice that, 
since pions and kaons are both Goldstone bosons of chiral symmetry breaking, 
one has in the chiral limit 
\be\label{App-Eq:H1perp-K-chiral-limit}
	\lim\limits_{m_K  \to 0}\frac{H_1^{\perp(1/2)a/  K}}{D_1^{a/  K}}=
	\lim\limits_{m_\pi\to 0}\frac{H_1^{\perp(1/2)a/\pi}}{D_1^{a/\pi}}\;.
\ee
This implies that in the real world with explicit chiral symmetry breaking,
i.e.\ for non-zero pion- and kaon-masses $m_\pi$ and $m_K$, 
one may assume the following relations to hold approximately
\ba
	\frac{H_1^{\perp(1/2)\bar s/K^+}}{D_1^{\bar s/K^+}}\approx
	\frac{H_1^{\perp(1/2)     u/K^+}}{D_1^{     u/K^+}}\approx 
	\frac{H_1^{\perp(1/2)   u/\pi^+}}{D_1^{   u/\pi^+}}\,,\nonumber\\
	\phantom{a}\nonumber\\
	\frac{H_1^{\perp(1/2){\rm unf}/  K^+}}{D_1^{{\rm unf}/  K^+}}\approx 
	\frac{H_1^{\perp(1/2){\rm unf}/\pi^+}}{D_1^{{\rm unf}/\pi^+}}
	\;,\label{App-Eq:H1perp-K}
\ea
where it is understood that the fragmentation of $d$- and $\bar u$-flavour 
into $K^+$ is unfavoured. The estimate (\ref{App-Eq:H1perp-K}) relies on the 
assumption that ``the way from the chiral limit to the real world situation''
proceeds quantitatively in a similar way for both polarization dependent and
independent quantities.
(Notice that the unpolarized ``favoured'' $\bar s$- and $u$-flavour 
fragmentations into $K^+$ are actually different ---
with the latter being smaller than the former \cite{Kretzer:2000yf}.
In the view of the precision of data, however, the effects of strangeness 
can be neglected due to the smallness of the corresponding pdfs.
E.g. the chiral quark soliton model predicts a negligible strangeness 
contribution to transversity (more precisely: to the tensor charge)
\cite{Kim:1996vk}.)

On the basis of (\ref{App-Eq:H1perp-K-chiral-limit}) we estimate
\ba\label{App-Eq:B-Gauss-H1perp12-Kaon-1}
   \la 2B_{\rm Gauss}H_1^{\perp(1/2)u/K^+}\ra &\approx&\;\;\;(1.0\pm 0.2)\% \;, \\
   \label{App-Eq:B-Gauss-H1perp12-Kaon-2}
   \la 2B_{\rm Gauss}H_1^{\perp(1/2){\rm unf}/K^+}\ra &\approx& -(1.0\pm 0.2)\% \;. 
\ea
From (\ref{App-Eq:B-Gauss-H1perp12-Kaon-1},~\ref{App-Eq:B-Gauss-H1perp12-Kaon-2})
we obtain after similar approximations as in App.~\ref{App:details}
the result in Fig.~\ref{Fig03:AUL2-x}f.


\end{document}